\documentclass[epj]{svjour}
\usepackage{graphicx}
\usepackage{xspace}
\begin{document}
%
%%%%% Personal Macros %%%%%%%%%%%%%%%%%%%
%\newcommand{\Al}{$^{26}$Al\ }
\newcommand{\about}{$\simeq$}
\newcommand{\degree}{$^{\circ}$}
\newcommand{\fluxrad}{ph~cm$^{-2}$s$^{-1}$rad$^{-1}$\ }
%%%%%%%%%%%%%%%%%%%%%%%%%%%%%%%%%%%%%%%%%
\newcommand{\Msol}{M\ensuremath{_\odot}\xspace}
\newcommand{\gam}{\ensuremath{\gamma}}
\newcommand{\flux}{ph~cm\ensuremath{^{-2}} s\ensuremath{^{-1}}\xspace}
\newcommand{\cms}{cm\ensuremath{^{-2}} s\ensuremath{^{-1}}\xspace}
\newcommand{\solar}{\ensuremath{_\odot}\xspace}
\newcommand{\Na}{\ensuremath{^{22}}Na\xspace}
\newcommand{\Al}{\ensuremath{^{26}}Al\xspace}
\newcommand{\Ti}{\ensuremath{^{44}}Ti\xspace}
\newcommand{\Sc}{\ensuremath{^{44}}Sc\xspace}
\newcommand{\Ca}{\ensuremath{^{44}}Ca\xspace}
\newcommand{\Mn}{\ensuremath{^{55}}Mn\xspace}
\newcommand{\Feff}{\ensuremath{^{55}}Fe\xspace}
\newcommand{\Coff}{\ensuremath{^{55}}Co\xspace}
\newcommand{\Ni}{\ensuremath{^{56}}Ni\xspace}
\newcommand{\Co}{\ensuremath{^{56}}Co\xspace}
\newcommand{\Fefsix}{\ensuremath{^{56}}Fe\xspace}
\newcommand{\Nifs}{\ensuremath{^{57}}Ni\xspace}
\newcommand{\Cofs}{\ensuremath{^{57}}Co\xspace}
\newcommand{\Fefs}{\ensuremath{^{57}}Fe\xspace}
\newcommand{\Fe}{\ensuremath{^{60}}Fe\xspace}
%%%%%%%%%%%%%%%%%%%%%%%%%%%%%%%%%%%%%%%%%%%%%%%%%%%
%Included for Gather Purpose only:
%input "E:\rod\science\papers\TeX_bib\bibtex\bib\misc\rod_references_inclCR.bib"

\title{Particle Acceleration in Cosmic Sites}
\subtitle{Astrophysics Issues in our Understanding of Cosmic Rays}
\author{Roland Diehl%\inst{1}
}                     % Do not remove
%\and Andrew W. Strong %\inst{1}
%}                     % Do not remove
%
%\offprints{rod}          % Insert a name or remove this line
%
\institute{Max Planck Institut f\"ur extraterrestrische Physik, Garching, Germany  (\email{rod@mpe.mpg.de})}
\date{Received: 30 Nov 2008 / Revised version: 20 May 2009}
\abstract{
Particles are accelerated in cosmic sites probably under conditions very different from those at terrestrial particle accelerator laboratories. Nevertheless, specific experiments which explore plasma conditions and stimulate particle acceleration carry significant potential to illuminate some aspects of the cosmic particle acceleration process. Here we summarize our current understanding of cosmic particle acceleration, as derived from observations of the properties of cosmic ray particles, and through astronomical signatures caused by these near their sources or throughout their journey in interstellar space. We discuss the candidate-source object variety, and what has been learned about their particle-acceleration characteristics. We conclude identifying open issues as they are discussed among astrophysicists. --
The cosmic-ray differential intensity spectrum across energies from 10$^{10}$eV to 10$^{21}$eV reveals a rather smooth power-law spectrum. Two kinks occur at the {\it knee} ($\simeq$10$^{15}$eV) and at the {\it ankle} ($\simeq$3~10$^{18}$eV). It is unclear if these kinks are related to boundaries between different dominating sources, or rather related to characteristics of cosmic-ray propagation. Currently we believe that Galactic sources dominate up to 10$^{17}$eV or even above, and the extragalactic origin of cosmic rays at highest energies merges rather smoothly with Galactic contributions throughout the 10$^{15}$--10$^{18}$eV range. Pulsars and supernova remnants are among the prime candidates for Galactic cosmic-ray production, while nuclei of active galaxies are considered best candidates to produce ultrahigh-energy cosmic rays of extragalactic origin.
The acceleration processes are probably related to shocks formed when matter is ejected into surrounding space from energetic sources such as supernova explosions or matter accreting onto black holes. Details of shock acceleration are complex, as relativistic particles modify the structure of the shock, and simple approximations or perturbation calculations are unsatisfactory. This is where laboratory plasma experiments are expected to contribute, to enlighten the non-linear processes which occur under such conditions.
\PACS{
      {96.40}{cosmic rays}   \and
      {95.30}{astrophysical plasma}
     } % end of PACS codes
} %end of abstract
\maketitle
\let\jnl=\rmfamily
\def\refe@jnl#1{{\jnl#1~}}%

\newcommand\aj{\refe@jnl{AJ}}%
          % Astronomical Journal
\newcommand\actaa{\refe@jnl{Acta Astron.}}%
  % Acta Astronomica
\newcommand\araa{\refe@jnl{ARA\&A}}%
          % Annual Review of Astron and Astrophys
\newcommand\apj{\refe@jnl{ApJ}}%
          % Astrophysical Journal
\newcommand\apjl{\refe@jnl{ApJ}}%
          % Astrophysical Journal, Letters
\newcommand\apjs{\refe@jnl{ApJS}}%
          % Astrophysical Journal, Supplement
\newcommand\ao{\refe@jnl{Appl.~Opt.}}%
          % Applied Optics
\newcommand\apss{\refe@jnl{Ap\&SS}}%
          % Astrophysics and Space Science
\newcommand\aap{\refe@jnl{A\&A}}%
          % Astronomy and Astrophysics
\newcommand\aapr{\refe@jnl{A\&A~Rev.}}%
          % Astronomy and Astrophysics Reviews
\newcommand\aaps{\refe@jnl{A\&AS}}%
          % Astronomy and Astrophysics, Supplement
\newcommand\azh{\refe@jnl{AZh}}%
          % Astronomicheskii Zhurnal
\newcommand\memras{\refe@jnl{MmRAS}}%
          % Memoirs of the RAS
\newcommand\mnras{\refe@jnl{MNRAS}}%
          % Monthly Notices of the RAS
\newcommand\na{\refe@jnl{New A}}%
  % New Astronomy
\newcommand\nar{\refe@jnl{New A Rev.}}%
  % New Astronomy Review
\newcommand\pra{\refe@jnl{Phys.~Rev.~A}}%
          % Physical Review A: General Physics
\newcommand\prb{\refe@jnl{Phys.~Rev.~B}}%
          % Physical Review B: Solid State
\newcommand\prc{\refe@jnl{Phys.~Rev.~C}}%
          % Physical Review C
\newcommand\prd{\refe@jnl{Phys.~Rev.~D}}%
          % Physical Review D
\newcommand\pre{\refe@jnl{Phys.~Rev.~E}}%
          % Physical Review E
\newcommand\prl{\refe@jnl{Phys.~Rev.~Lett.}}%
          % Physical Review Letters
\newcommand\pasa{\refe@jnl{PASA}}%
  % Publications of the Astron. Soc. of Australia
\newcommand\pasp{\refe@jnl{PASP}}%
          % Publications of the ASP
\newcommand\pasj{\refe@jnl{PASJ}}%
          % Publications of the ASJ
\newcommand\skytel{\refe@jnl{S\&T}}%
          % Sky and Telescope
\newcommand\solphys{\refe@jnl{Sol.~Phys.}}%
          % Solar Physics
\newcommand\sovast{\refe@jnl{Soviet~Ast.}}%
          % Soviet Astronomy
\newcommand\ssr{\refe@jnl{Space~Sci.~Rev.}}%
          % Space Science Reviews
\newcommand\nat{\refe@jnl{Nature}}%
          % Nature
\newcommand\iaucirc{\refe@jnl{IAU~Circ.}}%
          % IAU Cirulars
\newcommand\aplett{\refe@jnl{Astrophys.~Lett.}}%
          % Astrophysics Letters and Communications
\newcommand\apspr{\refe@jnl{Astrophys.~Space~Phys.~Res.}}%
          % Astrophysics Space Physics Research
\newcommand\nphysa{\refe@jnl{Nucl.~Phys.~A}}%
          % Nuclear Physics A
\newcommand\physrep{\refe@jnl{Phys.~Rep.}}%
          % Physics Reports
\newcommand\procspie{\refe@jnl{Proc.~SPIE}}%
          % Proceedings of the SPIE
\newcommand\jgr{\refe@jnl{J.Geoph.Res.}}%
          % Journal of geophysical research
\section{Introduction}
\label{intro}
\begin{figure}
\centering
\includegraphics[width=0.46\textwidth]{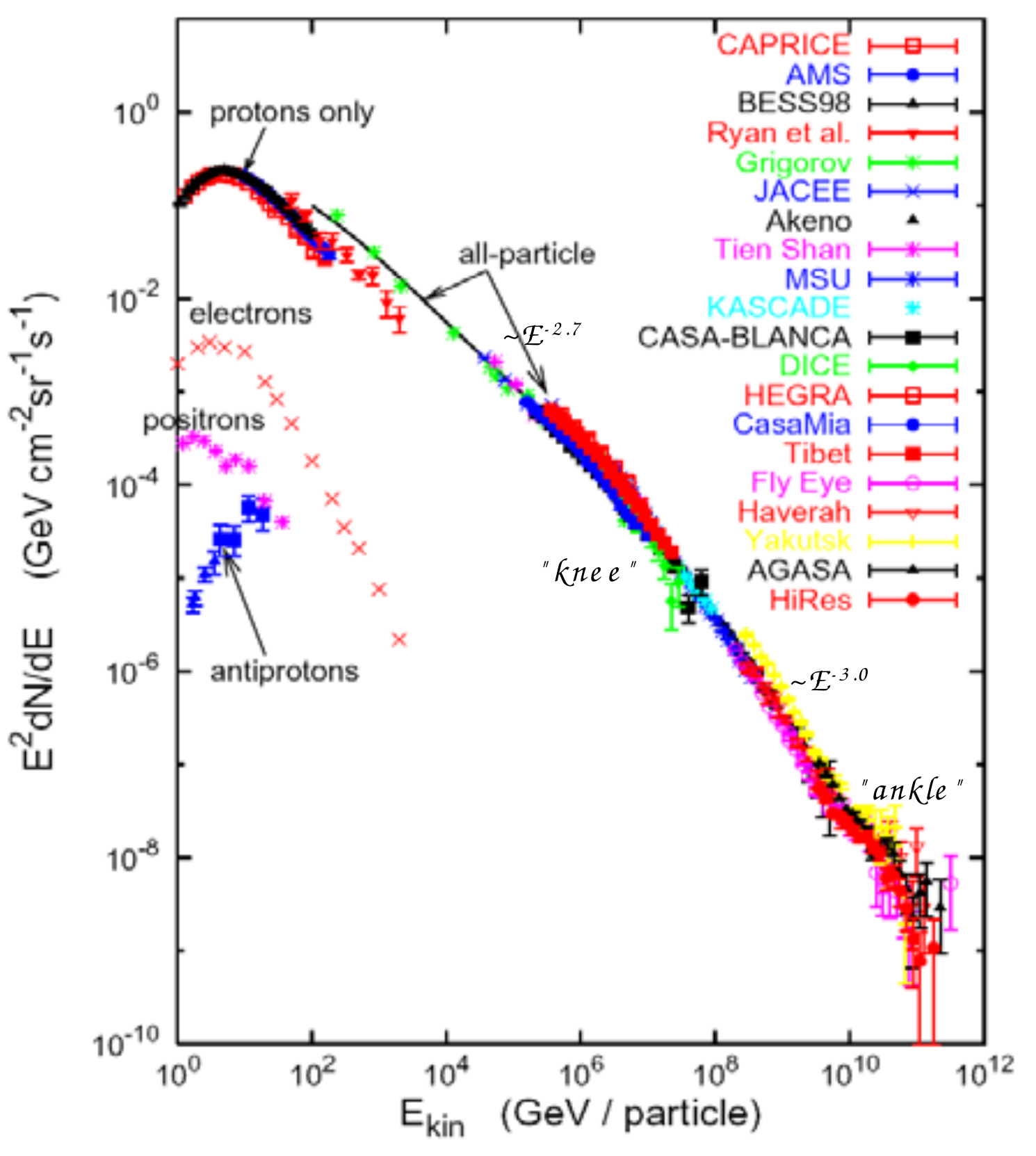}
\caption{The intensity spectrum of cosmic rays extends over more than ten orders of magnitude in energy with a surprisingly smooth spectrum $I(E)\sim E^{-2.7...3.0}$ (adapted from \cite{2006AIPC..861..630G}
%Gaisser 2006)
    }
\label{fig_CR-spectrum}
\end{figure}

Particle acceleration in laboratory environments mostly is based on generating an electric field to do the necessary acceleration work on charged particles. In cosmic sites, however, plasma currents would quickly short-circuit any eventually-present electrical fields. A magnetohydrodynamic view of cosmic plasma implies that no local electrical fields exist in the co-moving plasma reference frame, so no acceleration is obtained. How then is the strikingly-efficient acceleration of cosmic ray particles actually produced? Obviously, cosmic sites can provide environments for particle acceleration up to energies above 10$^{20}$eV \cite{2006JPhCS..47...15G} (see Fig.\ref{fig_CR-spectrum}). This is more than three orders of magnitude more than what we can achieve in terrestrial accelerator laboratories: The LHC at CERN will accelerate protons to energies up to 12 TeV in its final stage. A cosmic ray particle with an energy around 10$^{20}$eV (100~EeV) had been detected as early as 1962 \cite{1963PhRvL..10..146L}, %(Linsley, 1963)
more events were then seen above 10,000 TeV (=10~EeV) where fluxes are as low as one cosmic ray per square kilometer and year. As a field of astrophysics, research of relativistic-particle acceleration in cosmic sites had been initiated by Victor Hess, Albert Gockel, and Werner Kolh\"orster\cite{1912_ZPhys_Hess}: With their brave balloon measurements of ionizing radiation and its dependence with altitude in 1910-1913  up to 9~km altitude they demonstrated the existence of ionizing radiation impacting the Earth from outside. %(Hess 1912).
Combining and comparing astronomical findings about cosmic rays with particle acceleration studies in terrestrial laboratories is an essential part of our path to an understanding of the sources of cosmic rays. The extremely-high energies accessible only through cosmic rays provide a unique extension of the energy range being studied, albeit incurring the need for astrophysical modeling of the `experimental setup'.

It is the purpose of this paper to present our current understanding of cosmic particle acceleration and the nature of its sources, and to summarize the open issues herein, in order to cross-fertilize discussions among physicists on how particles can obtain relativistic energies, from the viewpoint of observational/experimental astrophysics.

%%%%%%%%%%%%%%%%%%%%%%%%%%%%%%%%%%%%%%%%%%%%%%%%%%%%%%%%%%%%%%%%%%%%%%%%%%%%%%%%%%%%%%%%
%
%%%%%%%%%%%%%%%%%%%%%%%%%%%%%%%%%%%%%%%%%%%%%%%%%%%%%%%%%%%%%%%%%%%%%%%%%%%
%%%%%%%%%%%%%%%%%%%%%%%%%%%%%%%%%%%%%%%%%%%%%%%%%%%%%%%%%%%%%%%%%%%%%%%%%%%

\section{Cosmic Rays}
\label{sec:cosmic_rays}

\subsection{Measuring Cosmic-Rays}
\label{sec:cosmic_rays_1}
After the recognition of {\it cosmic radiation} around 1910--1913, many experiments were set up to measure more details about cosmic rays in the upper atmosphere of the Earth and in near-Earth interplanetary space. From subsequent work of Robert Millikan, Arthur H. Compton, Pierre Auger, and others it became clear the ionizing radiation of extraterrestrial origins identified by the balloon experiments of Hess and others \cite{1912_ZPhys_Hess}\cite{1926AnP...384..572M}\cite{1939RvMP...11..232A}\cite{1939RvMP...11..288A}%(Hess 1912; Millikan 1926; Compton and Chou 1937; Auger et al 1939)
were caused by charged particles of extremely high energy and of possibly extragalactic origin, which created showers of secondary particles in the atmosphere. The discoveries of new elementary particles in cosmic ray showers, such as the muon, but also nucleonic excitation states, marks the decades of fruitful stimulation of nuclear physics and high-energy experiments by measurements on cosmic rays. Balloon experiments pioneered various techniques to record cosmic rays more directly than through their secondary showers, using emulsions and direct spectrometric analyzers with tracking chambers and magnets. Nowadays space experiments such as STARDUST \cite{1997M&PSA..32R..22B} %(Brownlee et al. 1997)
and AMS\cite{2008JPhCS.116a2001B} %(Battison 2008)
implement the technological forefront of those same technologies for capturing cosmic ray particles directly in interplanetary space. The  space missions of the Advanced Composition Explorer ACE (launched 1997) \cite{1998SSRv...86..285S} %Stone et al. 1998)
 and PAMELA \cite{2007APh....27..296P} (launched 2006) %(Picozza et al. 2007)
 provide detailed cosmic-ray intensity spectra discriminating different nuclear and lepton species (e.g. \cite{2005NuPhA.758..201I}), and balloon experiments such as HEAT (1994) and ATIC (2000, 2003) have a role specifically in searches for antimatter in cosmic rays \cite{1998ApJ...498..779B}\cite{2006AdSpR..37.1944P}\cite{2008Natur.456..362C}. %(Barwick et al. 1998; Panov et al. 2006, Chang et al. 2008).

Above energies of \about~10$^{14}$eV, particle fluxes become too low for such direct observations. Demands on detector volume and analyzing-magnet strength become excessive at energies exceeding thousands of GeV per nucleon. The technique of cosmic-ray studies then relies on using the Earth atmosphere directly as a detection volume, inferring the cosmic-ray particle characteristics from detailed measurements on the secondaries. % (e.g. Haungs et al. 2003).
Electromagnetic showers are initiated from the first interaction of an incident cosmic-ray particle at an altitude of 15 km and above. A variety of signatures can be recorded to re-constitute the nature (and direction) of the primary particle. Some of these techniques are only indirectly related to the primary particle, such as the widths or leptonic contents of showers, some are more direct such as Cerenkov radiation caused by particle motion at a velocity in excess of the local-medium's velocity of light. Nitrogen fluorescence light is emitted by air along trajectories of shower secondaries, and can be photographed from different viewing angles to construct a 3D shower image. The energy and flux of cosmic rays have been measured quite successfully with such techniques up to the extremes. Exciting results have been reported e.g. from the KASKADE ground level shower detector array of compositional changes around the {\it knee} of the cosmic ray spectrum \cite{2005APh....24....1A}. %(Antoni et al. 2005; see below).
But the nature of cosmic-ray particles is difficult to disentangle from the characteristics of the secondaries, as the first interaction with nuclei of the upper Earth atmosphere (or even interplanetary gas) breaks up nuclei into the individual nucleons and erases the primary composition. Extensive Monte-Carlo calculations of shower properties are being used (e.g. \cite{2003RPPh...66.1145H}), % Haungs et al. 2003),
but we must bear in mind that those calculations extrapolate high-energy physics interaction processes and cross sections significantly beyond where calibrations are feasible, and may not be realistic.

A combination of techniques has proven most successful for the current generation of cosmic-ray telescopes at highest energies: Secondary shower particles are recorded in surface detectors, and several telescopes capture an image of fluorescence light emitted along the trajectories of particles through the atmosphere. The AUGER experiment implements this technique over a large surface area of 3000~km$^2$ near Mendoza in Argentina \cite{1997SciAm.276a..32C}, %(Cronin et al. 1997),
and a corresponding site AUGER-North is under development in Colorado, USA.

\subsection{Indirect Observations about Cosmic Rays}
\label{sec:cosmic_rays_2}
Cosmic rays interact with matter and radiation along their trajectories through interstellar space. This produces characteristic {\it astronomical} signatures, which add unique insights into the nature of cosmic ray sources.

Secondary gamma-rays are readily observed at GeV energies, now also up to the TeV range, from interstellar gas, which generates a {\it source of photons} byh such cosmic-ray interactions through the processes of {\it Bremsstrahlung} and {\it inverse-Compton emission} as well as {\it hadronic interactions} \cite{2000ApJ...537..763S}\cite{2004ApJ...613..962S}. %(e.g. Strong et al. 2000; 2004).
Early measurements with SAS-II and COS-B established this aspect of Galactic cosmic-ray studies, which continued with the Compton Observatory and presently flourish from data captured with the instruments of the AGILE and FERMI missions. We distinguish several spectral regimes of the connection between cosmic rays and gamma-rays, depending on the characteristics of the productions of secondaries \cite{2007Ap&SS.309..465G}. %(Gabici and Aharonian 2007):
For ultrahigh primary cosmic rays in low-density surrounding gas with weak magnetic fields, {\it secondaries of the electromagnetic cascade} initiated by collision of a cosmic-ray particle with ambient plasma constituents are mainly produced in forward directions, and carry the same order of magnitude in energy as the primary cosmic rays. If the magnetic field is strong enough to significantly affect the motion of secondary electrons at the low-energy tail of the electromagnetic cascade, isotropic lepton clouds and thus a {\it pair-production halo} around the source would accompany the high-energy secondaries -- this halo, however, has not been measured yet. For even stronger magnetic fields, synchrotron energy losses result in a complete loss of the primary cosmic-ray's direction and energy, and lead to characteristic {\it synchrotron emission} caused by the high-energy cosmic rays \cite{2004A&A...419L..27B}. %(Berezhko and V\"olk 2004).
Observations of synchrotron emission extend from X-ray energies down into the radio regime, and are a valuable diagnostic to establish the existence of cosmic rays in specific source regions, due to the excellent astronomical precision of such observations. At energies below TeV, two major interactions dominate the secondary gamma-ray emission: The {\it creation of pions} (rest mass energy / threshold is 140~MeV in the collision rest frame) produces gamma-rays efficiently through the 2$\gamma$ decay of neutral pions $\pi^0$, and {\it Bremsstrahlung} is efficiently emitted from cosmic-ray collisions with ambient nuclei at energies below \about~GeV.

An additional process arises from the interaction of cosmic-ray electrons with ambient photons, up-scattering these in energy to contribute to the gamma-ray emissivity; this is the {\it inverse Compton scattering}, with a characteristic steep decline towards lower energies \cite{1997JApA...18...87P}\cite{2000thas.book.....P}. %(Padmanabhan 1997; 2000).
Photon fields for such interactions are provided from the cosmic microwave background as well as from stars and interstellar dust, but often more importantly from the cosmic-ray sources themselves through accretion disks or bright central sources and jets. Although understanding the inverse-Compton emission implies knowledge or assumptions about these photon fields, the characteristic shape of the inverse-Compton emission spectrum provides an important discrimination of leptonic versus hadronic origins of the secondary gamma-ray emission from cosmic rays.
Inverse-Compton energy losses upon ambient photons are relevant for cosmic-ray electrons (and positrons), and limit the distance which they can travel through interstellar space to $\sim$kpc, thus implying an origin within our Galaxy.

\subsection{Characteristics of Cosmic-Rays}
\label{sec:cosmic_rays_3}

The observed cosmic ray spectrum (Fig.~\ref{fig_CR-spectrum}) extends over ten orders of magnitude of energy with a remarkably smooth spectrum in the form of a powerlaw $I(E)=a\cdot E^{-\alpha}$. The powerlaw index $\alpha$ is $\simeq$2.7 at energies up to 10$^{15}$eV (the {\it knee}), then steepens towards $\alpha\simeq$3.0 with a small kink ({\it second knee}) near 10$^{17.5}$eV and $\alpha\simeq$3.3 beyond, before its slope flattens again near 3~10$^{18}$eV to $\alpha\simeq$3.0 at the {\it ankle}. Beyond, data become sparse, but another steepening appears above $\simeq 5\cdot 10^{19}$eV due to the {\it Greisen-Zatsepin-Kuz'min cut-off} (GZK) \cite{1966PhRvL..16..748G}\cite{1966ZhPmR...4..114Z}. %(Greisen 1966; Zatsepin and Kuzmin, 1966).
This cut-off is expected as the cosmic-ray protons above this energy exceed the pion production threshold when interacting with cosmic-microwave photons. A tentative detection of the GZK cutoff had been reported from the HiRes experiment \cite{2005PhLB..619..271A}, %(Abbasi et al. 2007),
and the AUGER results at highest energies \cite{2007Sci...318..938T} %(The Auger Collaboration, 2007)
are also consistent with this GZK cutoff, although the (uncertain) spatial distribution of cosmic ray sources and their inhomogeneities within the nearest few hundred Mpc also play a role for the spectral shape at the high end.

The composition of cosmic rays appears dominated by protons and atomic nuclei, electrons make up about 1--2\% of the total cosmic-ray flux at GeV energies, and even less above \cite{2006astro.ph..7109H}. %(Hillas 2006).
We must note that our knowledge about composition is best only at energies up to GeV energies; only a few pivotal composition measurements exist at much higher energies (see above). Positrons are observed at the level of 10$^{-3}$--10$^{-4}$ of the total, and anti-protons are two orders of magnitude lower in abundances still, at 1~GeV, yet with a rising spectrum so that their abundance becomes comparable to positrons at 10~GeV. Composition measurements at higher energies are a lively topic, reflected by the investments into balloon and space instrumentation, most notably the ambitious AMS project on the International Space Station ISS planning to place a large super-conducting magnet spectrometer in space \cite{2002NIMPA.478..119A}.
One of the main drivers here is the search for antimatter and for decay products of dark matter particles, both expected to appear at energies of 100~GeV and beyond. Reports about evidences for excess in fluxes of positrons \cite{2009PhRvL.102e1101A,2009Natur.458..607A} %(Adriani et al. 2008)
 and electrons \cite{2008Natur.456..362C} %(Chang et al. 2008)
 at such high energies have re-kindled discussions about dark-matter and SUSY signatures in cosmic ray spectra. The composition of elements and isotopes at sub-GeV energies are very detailed and precise, by contrast \cite{2005NuPhA.758..201I}\cite{2001AIPC..598..269W}\cite{2007SSRv..130..415W}. %(Israel et al. 2005;  Wiedenbeck et al. 2001 and 2007).
 Surprisingly, elemental abundances in cosmic rays are quite close to solar abundances over-all. With closer inspection, remarkable systematic differences appear, however: The light elements of charges 3-5 (Li, Be, B) are much more abundant in cosmic rays than in the solar system, also intermediate-mass elements below the iron peak are somewhat enriched in cosmic rays. This is understood from spallation reactions of energetic cosmic ray nuclei upon their collisions with ambient interstellar Hydrogen along their trajectories in interstellar space. Therefore, the abundances of all {\it secondary} isotopes produced by spallation reactions are enriched in cosmic rays with respect to the standard abundance distribution \cite{1970Ap&SS...6..377S}. %(Stecker 1970).
 The secondary-to-primary isotope abundances can be translated to an estimated average residence time of order 10$^7$~y, using nuclear spallation cross sections, an average density for interstellar matter, and cosmic-ray velocities \cite{1979ApJ...231..606S}. %(Stone and Wiedenbeck 1979).
 In particular several radioactive isotopes, which have been found in cosmic rays at low energies, most notably $^{10}$Be, $^{26}$Al, and $^{36}$Cl, determine such {\it cosmic-ray propagation ages} to 15$\pm$1.6~My \cite{2001ApJ...563..768Y}. % (Yanasak et al. 2001).

Those inferences, however, rely on a rather simple propagation model for cosmic-rays, the {\it leaky-box model}. In this model, cosmic rays are accelerated at their (unknown) sources, and diffuse through interstellar space as mediated by turbulent interstellar magnetic fields. At high energies, {\it leakage} of cosmic rays from the plane of the Galaxy into the halo is expected as magnetic rigidity increases. Although the leaky-box model is quite successful for illustrating several cosmic-ray properties such as the production of secondaries, cosmic-ray propagation is a complex superposition of many processes and boundary conditions, addressed best in numerical models (e.g. GALPROP,\cite{2000ApJ...537..763S}\cite{2004A&A...422L..47S}\cite{2007ARNPS..57..285S}). %Strong et al. 2000; 2004; 2007).
Current modeling accounts for 3-dimensional distributions of cosmic-ray sources, radiation fields, and interstellar gas. The interstellar medium of the Galaxy with its complex and rapidly-evolving morphology (see below) may plausibly lead to local and episodic deviations from such time-averaged modeling (e.g. \cite{2004A&A...422L..47S}). %Strong et al. 2004).

At higher energies, species abundances can only be determined more indirectly (see above).
The KASKADE experiment \cite{2005APh....24....1A} %(Antoni et al. 2005)
showed that He nuclei are the most abundant species near 10$^{15}$eV, and that rigidity cutoffs from interstellar magnetic fields suggest a dominance of heavier nuclei in the regime above the knee, in the transition region from Galactic origin of cosmic rays towards an extragalactic cosmic-ray component, which probably dominates above energies of 10$^{17}$eV. Results are somewhat confusing, as they suggest on one hand that the effective rigidity cutoff of cosmic rays around the {\it knee} (which is at lower energies for lighter nuclei) leads to a  heavier composition with increasing energy, on the other hand the reported Silicon abundances do not show any cutoff signature, and iron is only seen at highest energies and not at all at the lower energies where protons and helium dominate. Clearly, an assessment of systematic uncertainties is needed before conclusions should be drawn.

At energies below 10~GeV, the cosmic-ray spectrum is seen to bend down from the powerlaw shape describing the higher energies. This results from shielding by the magnetic fields of the Earth and Sun, deflecting cosmic rays due to their charge \cite{1973JGR....78.1502L}. %(Luhman and Earl 1973).
Additionally, secondary particles created by cosmic-ray interactions within the Earth atmosphere partly escape from the atmosphere, and then may be trapped in the Earth's magnetosphere. In particular, secondary neutrons streaming from the top of the atmosphere will decay into protons and abundantly be captured in radiation belts \cite{1974JGR....79.5159K}. % (Kanbach et al. 1974).
Solar-wind particles will provide an additional admixture to the low-energy spectrum of cosmic rays observed near Earth, so that inference of the extrasolar cosmic-ray contributions below 10~GeV/nucleon are uncertain.
Low-energy cosmic rays are trapped by the Earth's magnetic field and move along its field lines. Thus charged-particles penetrate deep into the atmosphere towards the Earth's magnetic poles. Arctic light is the observational evidence of the abundance of low-energy cosmic rays captured in those {\it radiation belts}, with {\it cutoff rigidity} depending on geomagnetic latitude.

This sensitivity to magnetic fields explains why an {\it astronomy with cosmic rays} is impossible: the curvature of the trajectory of a cosmic ray, as estimated from the {\it Larmor radius} (typical interstellar magnetic fields have strengths of a few $\mu$Gauss), even at a cosmic-ray energy of a PeV would be smaller than the distance to even the most nearby stars. So, except for the highest energies, direct astronomical traces back to their original sources are not possible from observations of cosmic rays.

Generally, cosmic rays are found to arrive near Earth with an isotropic distribution, at energies well above the modulation by magnetic fields of Earth and Sun. There had been some reports about deviations from isotropy, i.e. clustering of directions for highest energy cosmic rays from the AGASA experiment (see discussion by \cite{1999RvMPS..71..165C}). %Cronin, 1999).
More convincingly, for events above 40,000~TeV the AUGER experiment recently reported even an {\it astronomical} result: Their trajectories point back to nearby active galaxies (distributed along the {\it supergalactic plane} which describes matter distribution of the local universe on the scale of $\sim$100~Mpc), a result based on 81 cosmic-ray events \cite{2007Sci...318..938T}. %(Pierre Auger Collaboration, 2007).
%%%%%%%%%%%%%%%%%%%%%%%%%%%%%%%%%%%%%%%%%%%%%%%%%%%%%%%%%%%%%%%%%%%%%%%%%%%%%%%%%%%%%%%%%%%%%%%%%%%%%
%
%%%%%%%%%%%%%%%%%%%%%%%%%%%%%%%%%%%%%%%%%%%%%%%%%%%%%%%%%%%%%%%%%%%%%%%%%%%
\begin{figure}
\centering
\includegraphics[width=0.46\textwidth]{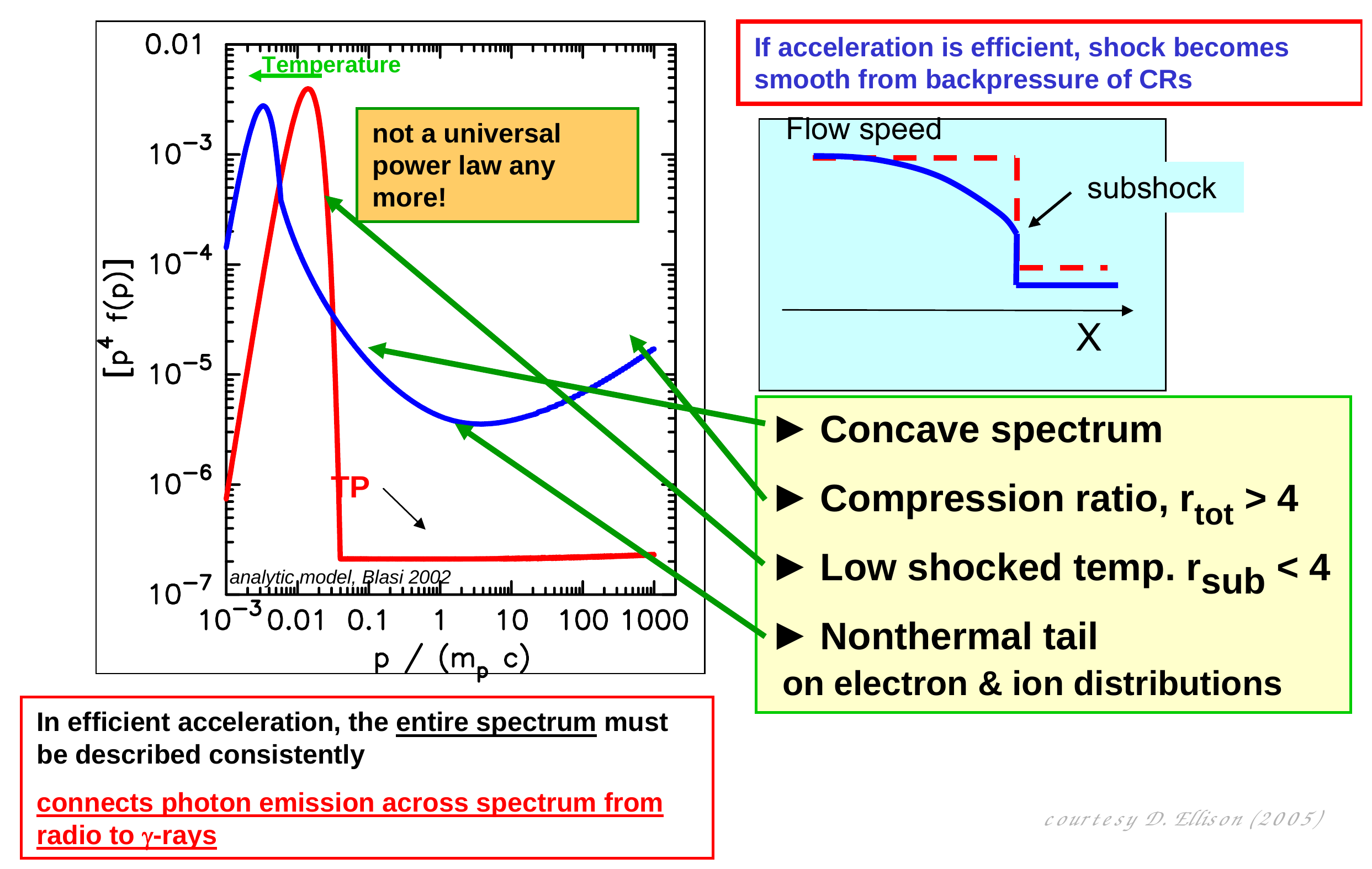}
\caption{Particle acceleration in an interstellar shock region: The feedback of the relativistic particle acceleration leads to substantial modifications to the structure of the shock in the theory of diffusive shock acceleration (DSA)  (figure courtesy Ellison, 2007)
    }
\label{fig_DSA}
\end{figure}
%%%%%%%%%%%%%%%%%%%%%%%%%%%%%%%%%%%%%%%%%%%%%%%%%%%%%%%%%%%%%%%%%%%%%%%%%%%

\section{Astrophysical Environments for Particle Acceleration}
\label{sec:astroplasma}
As we have seen above, energetic particles interact very efficiently with even low-density gas of the Earth's upper atmosphere. Hence, the energy for acceleration plausibly is provided either by macroscopic streaming or turbulence of interstellar gas, or within the energetic jet outflows initiated by violent accretion or explosion processes connected to compact objects and black holes. Interactions of energetic charged particles with such streaming gas or plasma produce what we observe as cosmic rays.

Astrophysical plasma conditions are characteristically different from the situation of an ideal gas (which often is in our minds when we attempt to understand cosmic gas).
A major difference is the predominance of Coulomb interactions with their long range, as opposed to short-range forces which characterize collisional equilibration ({\it thermalization}) in an ideal gas. This implies that cosmic plasma is inherently {\it viscous}, and the motions of electrons and ions are coupled over considerable ranges through electromagnetic field morphologies and dynamics; magnetic fields play a dominant role in mediating kinematic differences of cosmic particle populations. When collisions between individual charged particles even become negligible with respect to the electromagnetic field interactions, the plasma is called {\it collisionless} -- the standard case for astrophysical plasma \cite{2000thas.book.....P}\cite{2008ApJ...682L...5S}. %(Padmanabhan 2000; Spitkovsky 2008).

This leads to considerable complexities in treatments on smaller scales where energy transfers are substantial, such as in interstellar shocks (see below). On larger scales, it is often a useful approximation to consider the magnetic fields as {\it frozen} (or rather {\it coupled}) to the moving plasma. An assumed energy equilibration leads to the concept of a {\it multi-phase interstellar medium in pressure balance}, specifically cosmic-ray pressure balancing magnetic and interstellar gas dynamic pressures. Typical pressure values are $\sim$10$^{-12}$dyn~cm$^{-2}$ with a Galactic-plane magnetic field value near 2--5~$\mu$G. %   (e.g. Cox, 2006).

The more modern view of the interstellar medium is that of a dynamical and transitory medium, where energy injections by winds and supernova explosions are never balanced on scales below \about~100~pc (e.g. \cite{2005A&A...436..585D}). %deAvillez and Breitschwerdt 2005).
Therefore, in astrophysical plasma equilibria often are not obtained, hence perturbation approximations are inappropriate for treatment of dynamical processes such as acceleration. For example, the interactions between charged particles of very different masses, the electrons and ions, exchange very little energy, compared to interactions between particles of similar mass. Hence, {\it thermalization} occurs on different time scales for electrons and ions, and yet different between electrons and ions of a plasma. (For example, in a supernova remnant, electrons are characteristically thermalized within 100~y, while ion thermalization takes 4000~y, and electron-ion equilibrization needs 2~10$^5$~y \cite{2000thas.book.....P}; %(Padmanabhan, 2000);
supernova remnants are expected to efficiently accelerate particles while they are young and magnetic fields in the shock front are high due to the steepness of the shock, however.) This limits application of magnetohydrodynamic theory (MHD), which is based on the overall fluid properties of a plasma interacting with internal and external magnetic fields; mixtures of three different fluids with their respective properties add considerable complexity to such MHD descriptions. Note also that MHD theory is non-relativistic. Far from equilibria, plasma dynamics exhibits turbulence \cite{2000thas.book.....P}. In cosmic-source environments plausibly connected to relativistic acceleration, injection of plasma turbulence is expected, as one deals with cosmic explosions and formations of jets. The description of turbulent processes does not allow direct kinematic solution of plasma flows, it is largely restricted to following the energy transport between different spatial scales of turbulence, an {\it equilibrium turbulence} \cite{2000thas.book.....P}.

These considerations illustrate why processes of relativistic particle acceleration are commonly treated in {\it approximative models}. Acceleration in shocks thus was described in the {\it test particle approach}, where the cosmic-ray particle was {\it injected} into the externally-defined shock region with its density, magnetic-field, and velocity properties derived from hydrodynamical analogies (e.g. \cite{1994hea..book.....L}). %Longair 1994).
Calculation of the {\it Fermi acceleration process} then involves the determination of the mean free path of the cosmic-ray particle between {\it scatterings} on {\it magnetic-field turbulence} at the two different sides of the shock with their different bulk velocities ({\it momentum convection} process). Per scattering between the different sides of the shock, the energy gain is approximately $dE/E\approx 2dv/v$ (i.e., {\it Fermi-I} process, to first order in velocity difference across the shock; stochastic acceleration from random-motion scatterings ({\it momentum diffusion, Fermi-II-process}) would show a  quadratic relation). The energy flow from the shock region to the cosmic rays is ignored in this approach by definition. In this test particle model, the interaction zones which lead to acceleration through scattering of the test particle are set up independent of this acceleration process. Hydrodynamics leads us to expect temperature and density jumps across a shock front, which can be calculated analytically (e.g. density contrast by factor 4, from the Rankine-Hugoniot relation) \cite{1994hea..book.....L}. %(Longair 1994).

Refined theories of shock acceleration have been developed. Currently, the theory of {\it diffusive shock acceleration} (DSA) provides a description for supernova shocks, which reflects rather closely the observations of shock region radiation properties \cite{1999ApJ...526..385B}\cite{2001SSRv...99..305E}\cite{2005ApJ...632..920E}. % (Berezhko and Ellison 1999; Ellison 2001; Ellison et al. 2005).
In DSA, the feedback of relativistic-particle scattering on the overall plasma dynamics is included, leading to revised shock structure in temperature and density (e.g. \cite{2008ApJ...686..325L}). %Lee and Ellison 2008).
For example, the large, step-like temperature jump across the shock is reduced in magnitude, and smeared over a larger region on both sides of the shock. Conversely, the density changes across the shock become considerably stronger up to a factor eight. Since the magnetic-field strength is related to density, this incurs amplifications of the magnetic field which were absent in the test-particle treatment, but seem to occur in supernova remnants such as Cas A (see below). Note that this diffusive shock acceleration model also emphasizes a difference to typical laboratory situations, where magnetic diffusion may occur, while in astrophysical situations the magnetic field is frozen into the plasma and its high conductivity transfers diffusion processes to plasma currents only (cmp. \cite{2000thas.book.....P}). %Padmanabhan 2000).

After this general discussion, we now turn to specific cosmic environments where relativistic-particle acceleration is either plausible or has been seen already.
%%%%%%%%%%%%%%%%%%%%%%%%%%%%%%%%%%%%%%%%%%%%%%%%%%%%%%%%%%%%%%%%%%%%%%%%%%%%%%%%%%%%%%%%%%%%%%%%%%%%%

\section{Cosmic-Ray Sources}
\label{sec:sources}

From observations combined with astrophysical theory work, several candidate sites for production of relativistic charged particles have been identified. Although one or few of these source types may dominate, this is still an open field of research.
%
%%%%%%%%%%%%%%%%%%%%%%%%%%%%%%%%%%%%%%%%%%%%%%%%%%%%%%%%%%%%%%%%%%%%%%%%%%%
\begin{figure}
\centering
\includegraphics[width=0.46\textwidth]{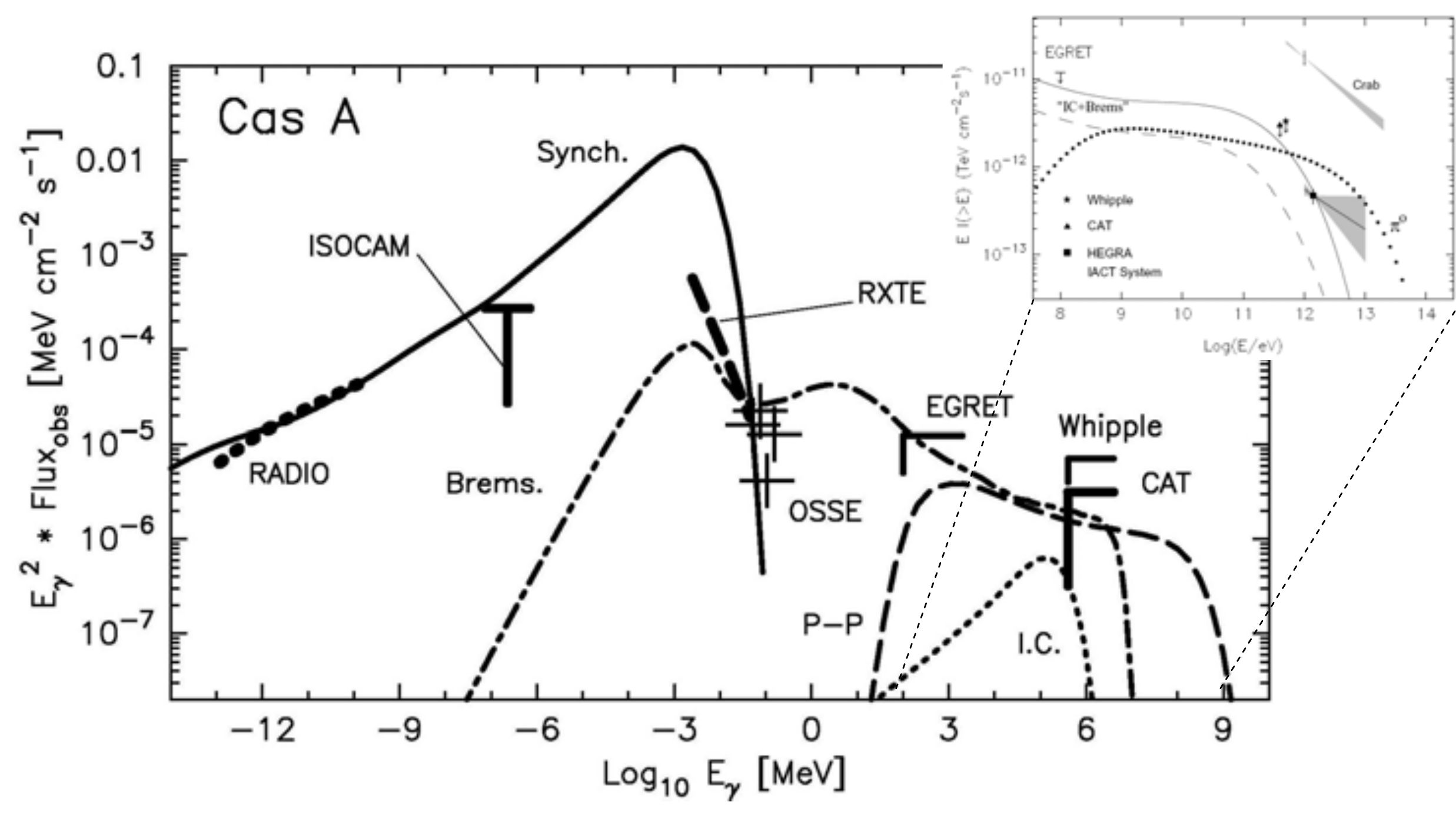}
\caption{Detailed models have been calculated for the emission of different types of electromagnetic radiation from a young supernova remnant accelerating particles. For the 350~year-old SNR Cas A at 3.4~kpc distance, most detailed observational constraints are available, for tesing those model predictions. This broad-band spectrum shows the different emission processes discussed in the text: (from low to high energies) Synchrotron emission is observable in a broad band from radio to X-ray energies; Bremsstrahlung occurs at hard-X and gamma-ray energies in the MeV to GeV band, pion-decay emission occurs in the GeV to TeV band as a broad flat-topped hump, similarly towards TeV energies the inverse-Compton emission has its peak.
(adapted from \cite{1999ApJ...526..385B}; see also  %Baring et al. 1999, Ellison et al. 2000, and Aharonian et al. 2004)
\cite{2000ApJ...540..292E,2004A&A...425L..13A})
    }
\label{fig_SNRCasA}
\end{figure}
%%%%%%%%%%%%%%%%%%%%%%%%%%%%%%%%%%%%%%%%%%%%%%%%%%%%%%%%%%%%%%%%%%%%%%%%%%%

\subsection{Supernova Remnants (SNR)}
\label{sec:sources:snrs}
The standard argument for supernova explosions as origin of the cosmic-rays arises from energy considerations: Propagation characteristics of cosmic rays imply that at highest energies they cannot be contained within the Galaxy, but escape. Observing a powerlaw spectrum and an assumed steady state demands that sources must replenish the energy lost from the Galaxy in the form of energetic particles. With an estimated confinement time (see above) of 10$^7$~years, and a local cosmic-ray energy density of \about~1.8~eV~cm$^{-3}$, one estimates an energy loss of about 2~10$^{41}$erg~s$^{-1}$. This energy loss is large by astrophysical standards, since most conventional cosmic objects have luminosities far below this. Supernovae occur at a rate ~1/(30--50 years), and eject a total kinetic energy on the order 10$^{51}$erg, hence generate a total energy input of 10$^{42}$erg~s$^{-1}$. Although an efficiency of 10\%  for the acceleration of particles appears rather high, in general this energy balance argument is adopted to make SNR the best candidate source of cosmic rays.
%
%%%%%%%%%%%%%%%%%%%%%%%%%%%%%%%%%%%%%%%%%%%%%%%%%%%%%%%%%%%%%%%%%%%%%%%%%%%
\begin{figure}
\centering
\includegraphics[width=0.46\textwidth]{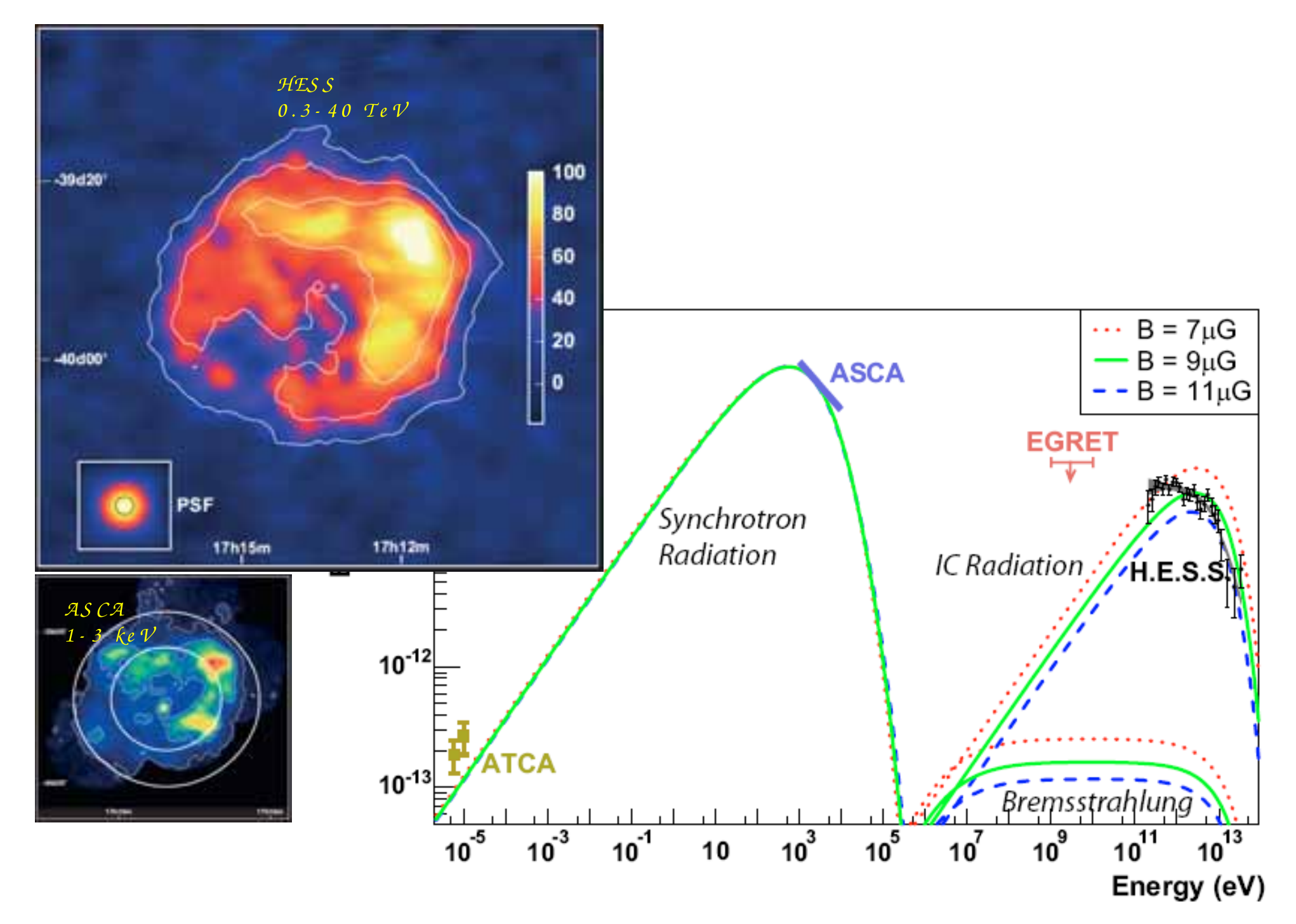}
\caption{Convincing evidence of relativistic-particle acceleration within a young supernova remnant has been accumulated for the SNR J1713.7-3946: X-ray emission attributed to electron synchrotron emission (small color image) appears spatially correlated with gamma-ray emission at TeV energies (larger image), the latter expected to arise either from inverse-Compton emission of the same electron population on ambient photons emitted by the SNR, or by pion decay from cosmic-ray protons accelerated together with the electrons. The broad-band spectrum in the right part of the Figure illustrates the different components of expected radiation (adapted from \cite{2006Natur.439..695A}) %Aharonian et al. 2006)
    }
\label{fig_SNR1713}
\end{figure}
%%%%%%%%%%%%%%%%%%%%%%%%%%%%%%%%%%%%%%%%%%%%%%%%%%%%%%%%%%%%%%%%%%%%%%%%%%%

The acceleration may plausibly occur in the shock resulting from the explosion into surrounding gas. The discontinuity between the exploding envelope traveling at several 1000~km~s$^{-1}$ and ambient interstellar gas at rest sets up a step change in density and temperature near this outward-propagating discontinuity. Charged particles may now cross the discontinuity, and may repeatedly scatter on either side to cross the shock again and again, thereby gaining net energy according to the Fermi-I acceleration mechanism. Such acceleration is only limited by increasing Larmor radii as the particle gains energy. Yet, acceleration depends on the efficiency of reflecting scatterings across the shock, hence on shock density, and on the velocity gradient across the shock. Therefore, supernova remnant acceleration of relativistic particles is intrinsically time-variable, depends on the age of the supernova remnant and its surrounding density structure. Young supernova remnants at ages 100--10000~years are best candidates, in general.

The {\it smoking gun} signifying that particle acceleration occurrs in SNR was believed to be the observation of synchrotron emission from relativistic particles in the magnetic field of the shock region, best coinciding with other clearly non-thermal radiation from high-energy particles such as Bremsstrahlung, pion-decay gamma-ray emission, or inverse-Compton emission \cite{1999ApJ...513..311B}. %(Baring et al. 1999).
SNR~1006 was a well-known X-ray source. When TeV gamma-rays were reported from the Cangaroo experiment \cite{1998ApJ...497L..25T}, %(Tanimori et al. 2000),
it was believed that this indeed was the {\it smoking gun}. Later more sensitive measurement by HESS demonstrated that the Cangaroo result was in error, however, and TeV emission would be much lower or even absent. Instead, SNR RXJ~1713.7-3946 is now considered a well-established proof of locally-accelerated particles within this young supernova remnant (see Fig.\ref{fig_SNR1713}; \cite{2004A&A...425L..13A,2005Ap&SS.300..255A,2007ApJ...665L..51A}. %Aharonian et al. 2004, 2006, 2007).
Here the morphology of the X-ray emission seen with the Japanese ASCA satellite instrument \cite{2003A&A...400..567U} %(Uchiyama et al. 2002)
follows closely the morphology of TeV emission as clearly mapped by the HESS telescopes. Unclear remains the origin of the TeV emission: If magnetic-field amplification within the shock should be large, synchrotron emission would be stronger and the corresponding electron population would be insufficient to explain the inverse-Compton emission. In that case, hadronic origin through pion decays from proton interactions in the SNR gas would be the explanation. But as the magnetic fields are uncertain, both explanations are possible.

\subsection{Stellar Wind Interactions}
\label{sec:sources:winds}
Most massive stars are known to occur in {\it binaries}, i.e., massive stars interact with other stars formed within the same parental molecular cloud, and capture collisions result in binary systems. Subsequent stellar evolution will lead to phases where stars have strong winds with mass losses on the order of 10$^{-4}$~\Msol~y$^{-1}$. For a subset of the binary systems, O-star winds may coincide with Wolf-Rayet wind phases of their companions. Depending on the orbital separations of such binaries, interacting stellar winds with typical wind velocities up to and exceeding 1000~km~s$^{-1}$ set up conditions very similar to what has been described above for supernova remnants. Therefore, such sources have also been discussed as candidate accelerators of cosmic-ray particles (e.g. \cite{2008AIPC.1085..157P}). %Paredes 2008).

\subsection{Pulsars}
\label{sec:sources:pulsars}
Pulsars are rapidly rotating neutron stars, with rotation periods as short as milliseconds for objects with a mass comparable to the mass of the Sun yet much more compact with typical radii of $\simeq$10~km. In general, the magnetic axis of the neutron star will not be aligned with the axis of rotation. The magnetic field of the pulsar will rotate with the star, but as it extends further out, an extreme condition arises at a radius where field lines corotating with the star would reach the speed of light. As a consequence, field lines not closing between the neutron star's magnetic poles within this {\it light-cylinder} will remain open. Plasma moving along these field lines will escape from the magnetosphere. Such loss of charges incurs plasma currents as well as regions devoid of charged particles further in along those open field lines. These devoid regions are called {\it gaps}, and are assumed to constitute the electrostatic acceleration regions for relativistic particles ejected by the pulsar. Radiation effects such as pair-production cascade emission and curvature radiation will produce X- and gamma-rays, relativistic aberration will add complexity and may be responsible for caustics (e.g. \cite{2004ApJ...606.1125D}.
%Dyks et al. 2004).

\begin{figure}
\centering
\includegraphics[width=0.46\textwidth]{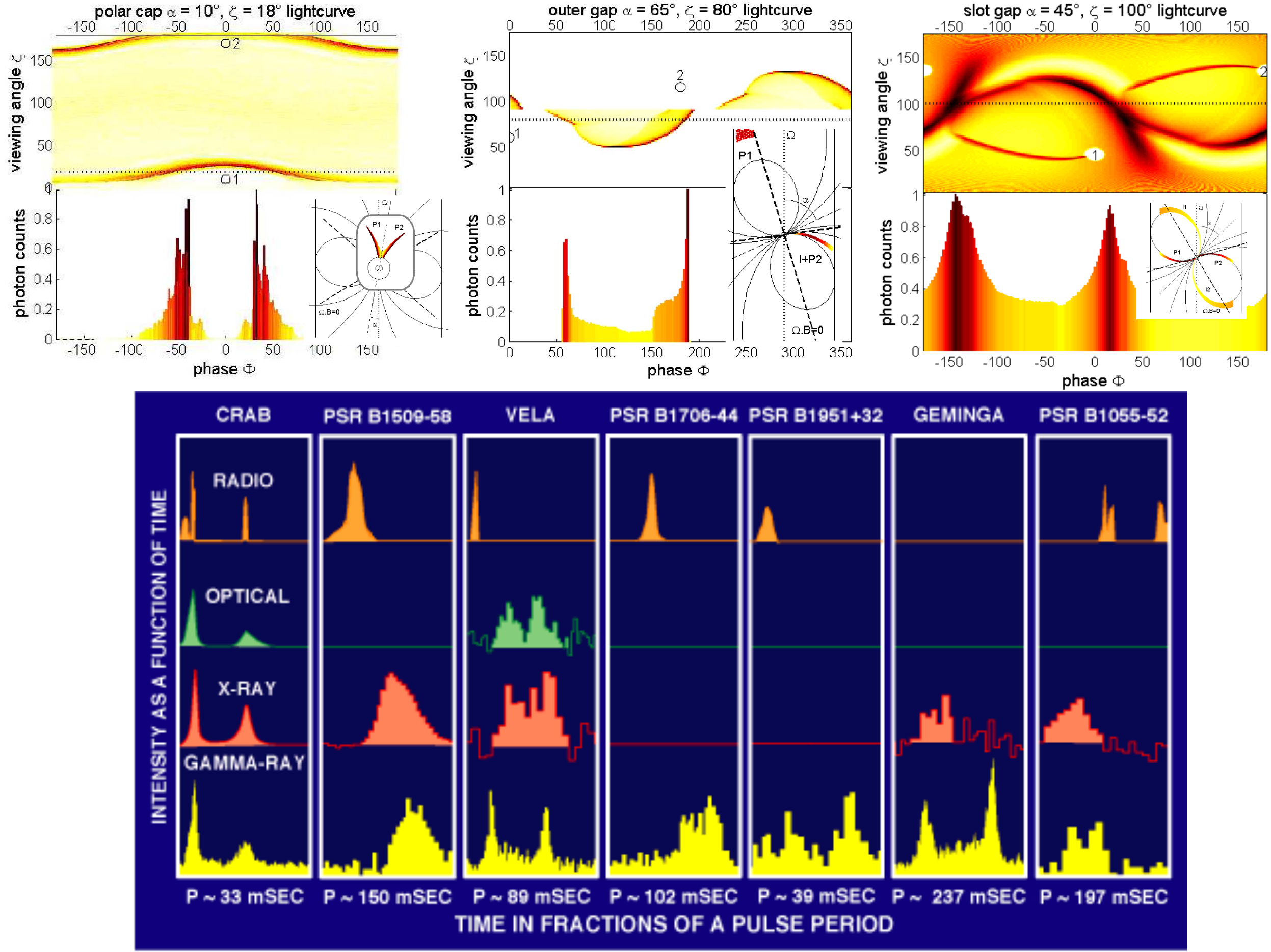}
\caption{Pulsar geometries are complex, the different aspect angles make it difficult to assemble a coherent and generalized pulsar emission model. Due to the rotation of the neutron star with its magnetic axis inclined, pair plasma generated along the inner field lines will flow out at open field line regions near the {\it light cylinder}. Voltage gaps are expected to result further in within the magnetosphere, which accelerate particles to relativistic energies up to possibly 10$^{18}$eV. It is unclear where exactly these acceleration {\it gaps} appear; the sketches show the gap viewing angles and light curves over the pulsar rotation phase for (from left to right) the {\it polar gap}, the {\it outer gap}', and the intermediate {\it slot gap} models, respectively. The lower part of the Figure shows observed light curves for seven pulsars, where emission is seen up to gamma-ray energies.
(adapted from \cite{2007AIPC..921...49H}\cite{2006NuPhA.777...98G}) %Harding 2006 and Harding \& Grenier 2006)
    }
\label{fig_pulsars}
\end{figure}

Aspect geometries may differ between different pulsars. Therefore, the association of peaks in the light curves over a pulsar rotation period cannot easily be assigned to either different poles of the rotating neutron star or different gaps sweeping over the observer's viewing direction. It is unclear where exactly these acceleration {\it gaps} appear: close to the neutron star surface would be the {\it polar gap}, closer to the light cylinder the {\it outer gap}, and  {\it slot gap} models describe intermediate cases, where the gaps would be very extended and located near the last closed field line. Fig.~\ref{fig_pulsars} shows those geometries, and the associated light curves expected under those viewing and gap-location cases. For comparison, the lower part of Fig.\ref{fig_pulsars} shows observed light curves for seven pulsars, from which detection of high-energy gamma-rays provided clearest evidence of currently-ongoing relativistic particle acceleration.

%
%%%%%%%%%%%%%%%%%%%%%%%%%%%%%%%%%%%%%%%%%%%%%%%%%%%%%%%%%%%%%%%%%%%%%%%%%%%
%%%%%%%%%%%%%%%%%%%%%%%%%%%%%%%%%%%%%%%%%%%%%%%%%%%%%%%%%%%%%%%%%%%%%%%%%%%

\subsection{Microquasars and Active Galaxies, and Gamma-Ray Bursts}
\label{sec:sources:jets}
Accretion onto compact stars and black holes is a well-known energy source and driver of high-energy emission of binaries at X- and gamma-ray energies. For a subclass of such binaries, the ejection of plasma jets with relativistic velocities was concluded from radio images taken at different epochs which showed apparent motion of radio emission knots at velocities in excess of the speed of light \cite{1994Natur.371...46M}. %(Mirabel et al. 1993).
More gravitational energy is available when the compact star is not a neutron star with a surface terminating inflow at a radius of 10--15~km, but rather a black hole. The Schwarzschild radius of a stellar object with the mass of the Sun is $\simeq$3~km, correspondingly smaller for more massive stellar remnants. Note that massive stars are created with a power-law distribution in masses extending out to about 80--100~\Msol\. Gravitational energy generated outside the Schwarzschild radius may escape the black hole and contribute to energy transfers taking part in the generation of these bi-polar plasma jets. About 30\% of X-ray binaries with radio emission were found to exhibit jets, and about eight micro-quasars are established to exist within our Galaxy. Clearly, ejection of plasma jets also constitutes a {\it smoking gun} of a relativistic-particle accelerator \cite{2005A&A...432..609B}. %(Bosch-Ramon et al. 2005).

Angular-momentum transfer in the flow of material from near such a compact object onto its surface results in formation of an accretion disk. Plasma flow within such an accretion disk may set up a dynamo, and thus be the driver of plasma jets perpendicular to the disk. Similar physical processes are expected, although at different scales, from accretion processes on supermassive black holes in the centers of active galactic nuclei \cite{2007ApJ...671...85N}. %(Neronov and Aharonian 2007).
Here, black-hole masses are on the order of 10$^8$~\Msol, and accretion is assumed to occur from nearby stars, being disrupted near the black hole. With correspondingly-larger Schwarzschild radii, temporal variations of the accretion luminosity will be slower and more difficult to trace observationally. Therefore, micro-quasars are the objects where astrophysicists strive to explore details of the physical processes at the heart of these accreting black holes.

Gamma-ray bursts of the {\it long} class (i.e., with burst durations in excess of 2~s) have been associated with {\it collapsars}, the gravitational collapses of massive stars to black holes at the end of their evolution \cite{2006ARA&A..44..507W}. %(Woosley and Bloom 2006).
The late stages of such a collapse suggest a similar situation, whereby the compact core of the massive star is accreted by the centrally-forming black hole. Again, formation of an accretion disk is expected, though with a lifetime on the order of milliseconds only, but the physical processes could be quite similar to what has been discussed in the context of jets formed in microquasars and active galaxies.

In theories of gamma-ray bursts, the internal-shock model had been proposed to explain the generation of gamma-ray emission from within the plasma jet \cite{1993ApJ...405..278M}. % (Meszaros and Rees 1993).
This model is quite successful in explaining the properties of the prompt gamma-ray emission and its variability. Plasma ejection in the jet with Lorentz factors of several hundred shift the $\gamma-\gamma$ pair production threshold to sufficiently-high energies, so that the accelerated-particle radiation may escape, as observed.

In principle, all these jet sources may accelerate particles to energies beyond the knee, thus being candidate sources for significant contributions to the observed cosmic-ray flux. AGN and GRB jet sources are powered by the largest energy reservoirs, and hence may be responsible for the ultrahigh cosmic rays which reach us from extragalactic sources.

\section{Open Issues}
\label{sec:issues}
Astrophysical first-order models are successful to explain much of the radiation observed from above candidate sources of cosmic rays. But some ad-hoc assumptions and first-order approximations are unsatisfactory, in particular the assumption of equilibria and the application of the perturbation method. Therefore, the connections derived between the source models the overall production of cosmic rays through such propagation models are highly uncertain. Derivations in particular of cosmic-ray fluxes and energy spectra, and of specific-source contributions, are vague.

Among the issues being discussed are:
\begin{itemize}
\item{}
What are the maximum-achievable energies for different candidate cosmic-ray accelerators?
Is there a transition between different source types?
\item{}
What are the acceleration efficiencies for different candidate accelerators?
Which magnetic-field amplifications occur in real shocks?
\item{}
How can the injection problem be solved?
Does self-injection happen? (Are non-thermal tails sufficient?)
\item{}
How critical is the geometry of shock regions?
Shocks are presently assumed to be planar. How does one properly treat non-planar shocks and their losses?
\item{}
Are observed GeV--TeV gamma-ray emissions leptonic or hadronic in origin?
What are the constraints on magnetic fields within the shock region, on densities, and on the radiation fields available for inverse-Compton scattering?
\end{itemize}
Obviously, the issues related to shock properties are all in some form accessible to laboratory experiments with plasma accelerated through laser pulses.
{\it Particle-in-cell} (PIC) codes for numerical treatments of the complex plasma processes with particle acceleration are now becoming more realistic with increased resolution in 3D, making use of advanced computing architectures \cite{2009..PRL_Henig}.
Still, {\it how} laboratory conditions can be set up to perform a sufficiently-controlled physics experiment, this remains a major challenge. Discussions such as experienced at this conference are steps in this direction.

{\bf Acknowledgements.} I am grateful for the conference organizers to stimulate discussions of particle acceleration across different fields of physics, and for inviting my contribution. I acknowledge in particular useful and stimulating discussions with Sergei Bulanov, Andrei Bykov, Don Ellison, Dieter Habs, Gottfried Kanbach, Andrew Strong, Toshiki Tajima and Peter Thirolf.

%
% BibTeX users please use
% \bibliographystyle{JHEP} %does NOT work well
%\bibliographystyle{aipproc} %does NOT work well
%% \bibliographystyle{elsart-num} % at least works
 \bibliographystyle{prsty}
% \bibliographystyle{elsart-harv} %does NOT work
%% \bibliography{rod_references_inclCR}

% \bibliography{cosmic-rays_rod,rod_references}

\begin{thebibliography}{10}

\bibitem{2006JPhCS..47...15G}
T.~K. {Gaisser}, Journal of Physics Conference Series {\bf 47},  15  (2006).

\bibitem{1963PhRvL..10..146L}
J. {Linsley}, Physical Review Letters {\bf 10},  146  (1963).

\bibitem{1912_ZPhys_Hess}
{Hess, V.}, {Z. Physik} {\bf {13}},  {1084}  ({1912}).

\bibitem{2006AIPC..861..630G}
I.~A. {Grenier} and A.~K. {Harding},  in {\em Albert Einstein Century
  International Conference}, Vol.~861 of {\em American Institute of Physics
  Conference Series} (American Institute of Physics, New York, 2006), pp.\
  630--637.

\bibitem{1926AnP...384..572M}
R.~A. {Millikan}, Annalen der Physik {\bf 384},  572  (1926).

\bibitem{1939RvMP...11..232A}
P. {Auger} and T. {Grivet}, Reviews of Modern Physics {\bf 11},  232  (1939).

\bibitem{1939RvMP...11..288A}
P. {Auger} {\it et~al.}, Reviews of Modern Physics {\bf 11},  288  (1939).

\bibitem{1997M&PSA..32R..22B}
D.~E. {Brownlee} {\it et~al.}, Meteoritics and Planetary Science, vol.~32, page
  A22 {\bf 32},  22  (1997).

\bibitem{2008JPhCS.116a2001B}
R. {Battiston}, Journal of Physics Conference Series {\bf 116},  012001
  (2008).

\bibitem{1998SSRv...86..285S}
E.~C. {Stone} {\it et~al.}, Space Science Reviews {\bf 86},  285  (1998).

\bibitem{2007APh....27..296P}
P. {Picozza} {\it et~al.}, Astroparticle Physics {\bf 27},  296  (2007).

\bibitem{2005NuPhA.758..201I}
M.~H. {Israel} {\it et~al.}, Nuclear Physics A {\bf 758},  201  (2005).

\bibitem{1998ApJ...498..779B}
S.~W. {Barwick} {\it et~al.}, \apj {\bf 498},  779  (1998).

\bibitem{2006AdSpR..37.1944P}
A.~D. {Panov} {\it et~al.}, Advances in Space Research {\bf 37},  1944  (2006).

\bibitem{2008Natur.456..362C}
J. {Chang} {\it et~al.}, \nat {\bf 456},  362  (2008).

\bibitem{2005APh....24....1A}
T. {Antoni} {\it et~al.}, Astroparticle Physics {\bf 24},  1  (2005).

\bibitem{2003RPPh...66.1145H}
A. {Haungs}, H. {Rebel}, and M. {Roth}, Reports on Progress in Physics {\bf
  66},  1145  (2003).

\bibitem{1997SciAm.276a..32C}
J.~W. {Cronin}, T.~K. {Gaisser}, and S.~P. {Swordy}, Scientific American {\bf
  276},  32  (1997).

\bibitem{2000ApJ...537..763S}
A.~W. {Strong}, I.~V. {Moskalenko}, and O. {Reimer}, \apj {\bf 537},  763
  (2000).

\bibitem{2004ApJ...613..962S}
A.~W. {Strong}, I.~V. {Moskalenko}, and O. {Reimer}, \apj {\bf 613},  962
  (2004).

\bibitem{2007Ap&SS.309..465G}
S. {Gabici} and F.~A. {Aharonian}, \apss {\bf 309},  465  (2007).

\bibitem{2004A&A...419L..27B}
E.~G. {Berezhko} and H.~J. {V{\"o}lk}, \aap {\bf 419},  L27  (2004).

\bibitem{1997JApA...18...87P}
T. {Padmanabhan}, Journal of Astrophysics and Astronomy {\bf 18},  87  (1997).

\bibitem{2000thas.book.....P}
T. {Padmanabhan}, {\em {Theoretical Astrophysics - Volume 1, Astrophysical
  Processes}} (Theoretical Astrophysics - Volume 1, Astrophysical Processes,
  ~Cambridge University Press, December 2000., Cambridge, 2000).

\bibitem{1966PhRvL..16..748G}
K. {Greisen}, Physical Review Letters {\bf 16},  748  (1966).

\bibitem{1966ZhPmR...4..114Z}
G.~T. {Zatsepin} and V.~A. {Kuz'min}, ZhETF {\bf 4},  114  (1966).

\bibitem{2005PhLB..619..271A}
R. {Abbasi} {\it et~al.}, Physics Letters B {\bf 619},  271  (2005).

\bibitem{2007Sci...318..938T}
{The Pierre Auger Collaboration}, Science {\bf 318},  938  (2007).

\bibitem{2006astro.ph..7109H}
A.~M. {Hillas}, ArXiv Astrophysics e-prints  (2006).

\bibitem{2002NIMPA.478..119A}
J. {Alcaraz} {\it et~al.}, Nuclear Instruments and Methods in Physics Research
  A {\bf 478},  119  (2002).

\bibitem{2009PhRvL.102e1101A}
O. {Adriani} {\it et~al.}, Physical Review Letters {\bf 102},  051101  (2009).

\bibitem{2009Natur.458..607A}
O. {Adriani} {\it et~al.}, \nat {\bf 458},  607  (2009).

\bibitem{2001AIPC..598..269W}
M.~E. {Wiedenbeck} {\it et~al.},  in {\em Joint SOHO/ACE workshop ''Solar and
  Galactic Composition''}, Vol.~598 of {\em American Institute of Physics
  Conference Series}, edited by R.~F. {Wimmer-Schweingruber} (American
  Institute of Physics, New York, 2001), p.\ 269.

\bibitem{2007SSRv..130..415W}
M.~E. {Wiedenbeck} {\it et~al.}, Space Science Reviews {\bf 130},  415  (2007).

\bibitem{1970Ap&SS...6..377S}
F.~W. {Stecker}, \apss {\bf 6},  377  (1970).

\bibitem{1979ApJ...231..606S}
E.~C. {Stone} and M.~E. {Wiedenbeck}, \apj {\bf 231},  606  (1979).

\bibitem{2001ApJ...563..768Y}
N.~E. {Yanasak} {\it et~al.}, \apj {\bf 563},  768  (2001).

\bibitem{2004A&A...422L..47S}
A.~W. {Strong} {\it et~al.}, \aap {\bf 422},  L47  (2004).

\bibitem{2007ARNPS..57..285S}
A.~W. {Strong}, I.~V. {Moskalenko}, and V.~S. {Ptuskin}, Annual Review of
  Nuclear and Particle Science {\bf 57},  285  (2007).

\bibitem{1973JGR....78.1502L}
J.~G. {Luhmann} and J.~A. {Earl}, \jgr {\bf 78},  1502  (1973).

\bibitem{1974JGR....79.5159K}
G. {Kanbach}, C. {Reppin}, and V. {Schoenfelder}, \jgr {\bf 79},  5159  (1974).

\bibitem{1999RvMPS..71..165C}
J.~W. {Cronin}, Reviews of Modern Physics Supplement {\bf 71},  165  (1999).

\bibitem{2008ApJ...682L...5S}
A. {Spitkovsky}, \apjl {\bf 682},  L5  (2008).

\bibitem{2005A&A...436..585D}
M.~A. {de Avillez} and D. {Breitschwerdt}, \aap {\bf 436},  585  (2005).

\bibitem{1994hea..book.....L}
M.~S. {Longair}, {\em {High energy astrophysics. Vol.2: Stars, the galaxy and
  the interstellar medium}} (Cambridge: Cambridge University Press, 1994, 2nd
  ed., Cambridge, 1994).

\bibitem{1999ApJ...526..385B}
E.~G. {Berezhko} and D.~C. {Ellison}, \apj {\bf 526},  385  (1999).

\bibitem{2001SSRv...99..305E}
D.~C. {Ellison}, Space Science Reviews {\bf 99},  305  (2001).

\bibitem{2005ApJ...632..920E}
D.~C. {Ellison} and G. {Cassam-Chena{\"i}}, \apj {\bf 632},  920  (2005).

\bibitem{2008ApJ...686..325L}
S.-H. {Lee}, T. {Kamae}, and D.~C. {Ellison}, \apj {\bf 686},  325  (2008).

\bibitem{2000ApJ...540..292E}
D.~C. {Ellison}, E.~G. {Berezhko}, and M.~G. {Baring}, \apj {\bf 540},  292
  (2000).

\bibitem{2004A&A...425L..13A}
F. {Aharonian} {\it et~al.}, \aap {\bf 425},  L13  (2004).

\bibitem{2006Natur.439..695A}
F. {Aharonian} {\it et~al.}, \nat {\bf 439},  695  (2006).

\bibitem{1999ApJ...513..311B}
M.~G. {Baring} {\it et~al.}, \apj {\bf 513},  311  (1999).

\bibitem{1998ApJ...497L..25T}
T. {Tanimori} {\it et~al.}, \apjl {\bf 497},  L25+  (1998).

\bibitem{2005Ap&SS.300..255A}
F. {Aharonian} and A. {Neronov}, \apss {\bf 300},  255  (2005).

\bibitem{2007ApJ...665L..51A}
J. {Albert} {\it et~al.}, \apjl {\bf 665},  L51  (2007).

\bibitem{2003A&A...400..567U}
Y. {Uchiyama}, F.~A. {Aharonian}, and T. {Takahashi}, \aap {\bf 400},  567
  (2003).

\bibitem{2008AIPC.1085..157P}
J.~M. {Paredes},  in {\em American Institute of Physics Conference Series},
  Vol.~1085 of {\em American Institute of Physics Conference Series}, edited by
  F.~A. {Aharonian}, W. {Hofmann}, and F. {Rieger} (American Institute of
  Physics, New York, 2008), pp.\ 157--168.

\bibitem{2004ApJ...606.1125D}
J. {Dyks}, A.~K. {Harding}, and B. {Rudak}, \apj {\bf 606},  1125  (2004).

\bibitem{2007AIPC..921...49H}
A.~K. {Harding},  in {\em The First GLAST Symposium}, Vol.~921 of {\em American
  Institute of Physics Conference Series}, edited by S. {Ritz}, P. {Michelson},
  and C.~A. {Meegan} (American Institute of Physics, New York, 2007), pp.\
  49--53.

\bibitem{2006NuPhA.777...98G}
T.~K. {Gaisser} and T. {Stanev}, Nuclear Physics A {\bf 777},  98  (2006).

\bibitem{1994Natur.371...46M}
I.~F. {Mirabel} and L.~F. {Rodriguez}, \nat {\bf 371},  46  (1994).

\bibitem{2005A&A...432..609B}
V. {Bosch-Ramon}, F.~A. {Aharonian}, and J.~M. {Paredes}, \aap {\bf 432},  609
  (2005).

\bibitem{2007ApJ...671...85N}
A. {Neronov} and F.~A. {Aharonian}, \apj {\bf 671},  85  (2007).

\bibitem{2006ARA&A..44..507W}
S.~E. {Woosley} and J.~S. {Bloom}, \araa {\bf 44},  507  (2006).

\bibitem{1993ApJ...405..278M}
P. {Meszaros} and M.~J. {Rees}, \apj {\bf 405},  278  (1993).

\bibitem{2009..PRL_Henig}
A. {Henig} {\it et~al.}, PRL {\bf {in press}},    (2009).

\end{thebibliography}
%
%Linsley J, Phys Rev Lett 10, 146 (1963) % first detection of 10^20 eV Crs
%Pierre Auger Collaboration, Science 318, 938 (2007) % Crs from AGN?
%Israel M.H. et al., NuclPhys A 758, 201c (2005) % CR composition
% Panov A.D. et al., Adv Sp Res 37, 1944 (2006) % ATIC CR nuclei measurement
% Chang J. et al., Nature 456, 362 (2008) % positron excess
% Haungs A et al., Rep Prog Part Phys 66, 1145 (2003) % CR shower properties
% Antoni T. et al., Astropart.Phys 24, 1 (2005) % CR composition at knee
% Cronin J.W., Sci Am 276, 46 (1997) % AUGER concept
%\begin{thebibliography}{}
% etc
%\end{thebibliography}

\end{document}